\documentclass[%
 reprint,
 amsmath,amssymb,
 aps,
floatfix,
]{revtex4-2}

\usepackage{graphicx}
\usepackage{dcolumn}
\usepackage{bm}

\begin{document}

\preprint{APS/123-QED}

\title{The effect of thermal photons on exceptional points in coupled resonators}

\author{Grzegorz Chimczak}
 \email{chimczak@amu.edu.pl}
 \author{Anna Kowalewska-Kud{\l}aszyk}%
 \author{Ewelina Lange}%
 \author{Karol Bartkiewicz}%
 \altaffiliation[Also at ]{RCPTM, Joint Laboratory of Optics of Palacký University and Institute of Physics of Czech Academy of Sciences, 17. listopadu 12, 771 46 Olomouc, Czech Republic}
 \affiliation{Institute of Spintronics and Quantum Information, Faculty of Physics, Adam Mickiewicz University, 61-614 Pozna\'{n}, Poland}%
\author{Jan Pe\v{r}ina Jr.}
\affiliation{RCPTM, Joint Laboratory of Optics of Palacký University and Institute of Physics of Czech Academy of Sciences, 17. listopadu 12, 771 46 Olomouc, Czech Republic
}%

\begin{abstract}
We analyse two quantum systems with hidden parity-time (${\cal{PT}}$) symmetry: one is an optical device, whereas another is a superconducting microwave-frequency device. To investigate their symmetry, we introduce an equilibrium frame, in which loss and gain terms for a given Hamiltonian are balanced. We show that the non-Hermitian Hamiltonians of both systems can be tuned to reach an exceptional point (EP), i.e., the point in parameter space at which a transition from broken to unbroken hidden ${\cal{PT}}$ symmetry takes place. We calculate a degeneracy of a Liouvillian superoperator, which is called the Liouvillian exceptional point (LEP), and show that, in the optical domain, LEP is equivalent to EP obtained from the non-Hermitian Hamiltonian (HEP). We also report breaking the equivalence between LEP and HEP by a non-zero number of thermal photons for the microwave-frequency system.
\end{abstract}

\maketitle

\section{Introduction}
In recent years, there has been increasing interest in exploring non-Hermitian systems as a source of novel physical effects [for examples see Refs.~\citenum{ozdemir2019parity,el2018non,Miri2019,Ashida20}]. It has been shown that the special group of non-Hermitian Hamiltonians, which is parity-time (${\cal{PT}}$) symmetric, exhibits entirely real spectra, like Hermitian Hamiltonians, in the region of a parametric space where this symmetry is in unbroken phase~\cite{Bender_98,bender2007making,BenderBook}. From both a theoretical and an experimental points of view, much more interesting than non-Hermitian Hamiltonians having entirely real spectra are degeneracies of these Hamiltonians, which are placed in points of the parameter space, where a phase transition occurs from an unbroken to a broken ${\cal{PT}}$ symmetry. Such degeneracies, known as exceptional points (EPs), are the points of the parametric space where the eigenvalues and the corresponding eigenvectors of a Hamiltonian coincide~\cite{Kato66}. Only non-Hermitian Hamiltonians can display EPs~\cite{Mostafazadeh15}, and therefore, only in non-Hermitian systems, all the interesting physics associated with these degeneracies can be observed. These nontrivial phenomena include enhancement of weak signal sensing \cite{Chen_17}, enhancement of spontaneous emission \cite{Lin_16}, asymmetric light propagation \cite{Markis_08,Jin17}, single-mode laser \cite{Feng_14}, electromagnetically induced transparency \cite{Wang_20} just to name the few. 
The EPs are usually studied in the semiclassical regime, where optical and photonic systems are driven with strong classical external fields. Recently, these studies have been extended to the fully quantum regime~\cite{Hatano19, Perina_19, Minganti_19}. These studies do not use the Schr\"odinger equation with a non-Hermitian Hamiltonian, but are based on two different fully quantum descriptions of the open system dynamics, namely the master equation with a Liouvillian superoperator and the Heisenberg-Langevin equations. Since the matrix form of a Liouvillian superoperator is a non-Hermitian matrix, it can display degeneracies just like non-Hermitian Hamiltonians in the semiclassical regime. These degeneracies, known as Liouvillian EPs (LEPs), and their influence on features of the quantum system are attracting increasing attention~\cite{Minganti18, Arkhipov_20a}. The master equation approach allows investigating fully quantum exceptional points, i.e., LEPs. It also helps to design a quantum system associated with a given non-Hermitian Hamiltonian. The quantum dynamics included in the master equation can be decomposed to give quantum trajectories~\cite{carmichael_traj,PlenioKnight_traj}. In the quantum trajectory method, the evolution of an open quantum system during the time intervals without quantum jumps is governed by a non-Hermitian Hamiltonian. Therefore, it is possible to realise non-Hermitian Hamiltonians in the fully quantum system using postselection. In this way, EP of a non-Hermitian Hamiltonian (HEP) has recently been observed in an experimental superconducting system~\cite{Naghiloo19}. The important difference between LEP and HEP is in accounting for quantum jumps. The former includes quantum jumps, whereas the latter assumes their absence. Therefore, in general, LEPs are different from HEPs. Shortly after the first observation of HEP in a fully quantum system, several papers were published comparing LEPs with HEPs~\cite{Arkhipov_20b, Wiersig20, Minganti_20, Ou2021,Chen21}. It was shown that in some systems LEP can be equivalent to HEP in the sense that the position of both in the parameter space is the same. In one of the mentioned papers, Arkhipov \emph{et al.}~\cite{Arkhipov_20b} have investigated a quantum system composed of two coupled cavities, where one cavity experiences incoherent gain, while another only damping, and have found such equivalence of LEP and HEP. 

In the present work, we study a quantum system consisting of two laser-driven coupled optical cavities, from which a field leaks out to the reservoir. The non-Hermitian Hamiltonian describing this system is not ${\cal{PT}}$-symmetric, because it does not include an incoherent gain term. Nevertheless, we find the position of the HEP by revealing the ${\cal{PT}}$ symmetry hidden in this Hamiltonian and the point where a phase transition occurs. To this end, we introduce the idea of the equilibrium frame (EF) --- a frame, where the hidden ${\cal{PT}}$ symmetry is clearly seen. We demonstrate that in a system with coherent gain HEP is equivalent to LEP. A similar observation was reported for a system with incoherent gain in Ref.~\citenum{Arkhipov_20b}. Finally, we consider a superconducting circuit realised in the microwave domain, which is described by the same master equation as the optical system in the case when the thermal photon number in a thermal environment is negligible. We report breaking the equivalence between LEP and HEP by a non-zero number of thermal photons.

\section{Results}
\subsection{Equilibrium frame}
The main idea of the transformation to equilibrium frame (EF) is based on a rotating frame transformation, frequently used in quantum optics. We assume that the total Hamiltonian can be written as a sum of two terms $H^{\cal{PT}}$ and $H_{0}$. The Schr\"odinger equation is thus given by ($\hbar=1$)
\begin{eqnarray}
  \label{eq:EF01}
  i\partial_t|\psi\rangle &=& ( H^{\cal{PT}} + H_{0} ) |\psi\rangle\, .
\end{eqnarray}
Now we make the substitution $|\psi\rangle=S\, |\widetilde{\psi}\rangle$, where $S$ and $|\widetilde{\psi}\rangle$ are time-dependent. If we set $S=\exp(-i\, H_{0}\, t)$ then the Schr\"odinger equation reduces to 
\begin{eqnarray}
  \label{eq:EF02}
  i\partial_t |\widetilde{\psi}\rangle &=& \widetilde{H} |\widetilde{\psi}\rangle\, ,
\end{eqnarray}
where $\widetilde{H} = S^{-1} H^{\cal{PT}} S$. In the case of the transformation to a rotating frame, $S$ is unitary, because $H_{0}$ is Hermitian. However, in the transformation to EF the operator $S$ is not a unitary one, because $H_{0}$ is not Hermitian. In both cases $\widetilde{H}$ and $H^{\cal{PT}}$ have the same eigenvalues. In order to obtain a ${\cal{PT}}$-symmetric Hamiltonian in EF, we restrict ourselves to the cases, where $[H^{\cal{PT}},H_{0}]=0$. Using the Baker–Hausdorff lemma
\begin{eqnarray}
  \label{eq:EF03}
  e^Y X e^{-Y} &=& X+[Y,X]+(1/2!)\big[Y,[Y,X]\big]+\dots
\end{eqnarray}
one can easily prove that $\widetilde{H}=H^{\cal{PT}}$ for these cases.

For $[H^{\cal{PT}},H_{0}]=0$, both Hamiltonians have the same set of eigenstates, and then we may relate the eigenvalues of the Hamiltonian given in the initial frame (IF) to those in EF. Therefore, an $i$-th eigenvalue of the total Hamiltonian in IF
\begin{eqnarray}
  \label{eq:EF04}
  H |\phi_{i}\rangle &=&  H^{\cal{PT}} |\phi_{i}\rangle + H_{0} |\phi_{i}\rangle 
  = \widetilde{H} |\phi_{i}\rangle + H_{0} |\phi_{i}\rangle \nonumber\\
  E_{i} |\phi_{i}\rangle &=& \widetilde{E}_{i} |\phi_{i}\rangle + E^{(0)}_{i} |\phi_{i}\rangle
\end{eqnarray}
 is equal to the sum of the $i$-th eigenvalue of $\widetilde{H}$ and the corresponding eigenvalue of $H_{0}$. This fact is important when one is looking for Hamiltonians displaying EPs, i.e., points in the parameter space, where two (or more) eigenvalues have the same value. If $H^{\cal{PT}}$ is PT-symmetric, then the Hamiltonian in EF, i.e., $\widetilde{H}$, can display EPs. In such a case, at least two eigenvalues of $\widetilde{H}$ coincide (  $\widetilde{E}_{i}=\widetilde{E}_{j}$ for some $i$ and $j$). 
 Therefore, one may state that the Hamiltonian given in IF, i.e., $H=H^{\cal{PT}}+H_{0}$, being
 not a PT-symmetric one, can also display EP if $E^{(0)}_{i}=E^{(0)}_{j}$. 
 The second condition for EP is also fulfilled because $H^{\cal{PT}}$ and $H$ have the same set of eigenstates. Therefore, in this point the eigenvectors of $H$ also coincide. Thus, we can say that EF reveals the hidden symmetry of $H$. 

The existence of degenerate eigenvalues of $H_{0}$ is a necessary condition for $H$ displaying EP. Moreover, the eigenvalues suggest the convenient frame. If the eigenvalues of $H_{0}$ are real, then we transform to a rotating frame. If they are imaginary, we deal with a transformation to a frame in which the length of the eigenstates scales with time. In the case when a system is in an unbroken, ${\cal{PT}}$-symmetric phase, i.e., $H^{\cal{PT}}$ has a real spectrum and the eigenvalues of $H_{0}$ are imaginary, we can associate these two parts of $H$ with the observable  energy of the system ($H^{\cal{PT}}$) and the metric describing the geometric nature of the Hilbert space ($H_{0}$)~\cite{Zhang19a,Mostafazadeh18,Zhang19b}.

It should be noted that condition $[H^{\cal{PT}},H_{0}]=0$ does not mean that $H_{0}$ is a constant of motion, since $H^{\cal{PT}}$ is not Hermitian. The conserved quantities for ${\cal{PT}}$-symmetric Hamiltonian are given by intertwining operators~\cite{Bian20_c,MOSTAFAZADEH10_pH}.

It is also worth mentioning that the equilibrium frame can also be useful to reveal hidden pseudo-Hermiticity of non-Hermitian Hamiltonians. It is known that the ${\cal{PT}}$ symmetry is a special case of pseudo-Hermiticity~\cite{Mostafazadeh02_1,Mostafazadeh02_5,Mostafazadeh02_8}. If the total Hamiltonian can be written as a sum of commuting parts, i.e., a pseudo-Hermitian Hamiltonian and $H_{0}$, then one can expect that the eigenvalues of the Hamiltonian given in IF are related to those in EF.

\subsection{Hidden ${\cal{PT}}$ Symmetry of Passive Optical System with Coherent Gain}
\begin{figure}[htbp]
\centering
\fbox{\includegraphics[width=0.6\linewidth]{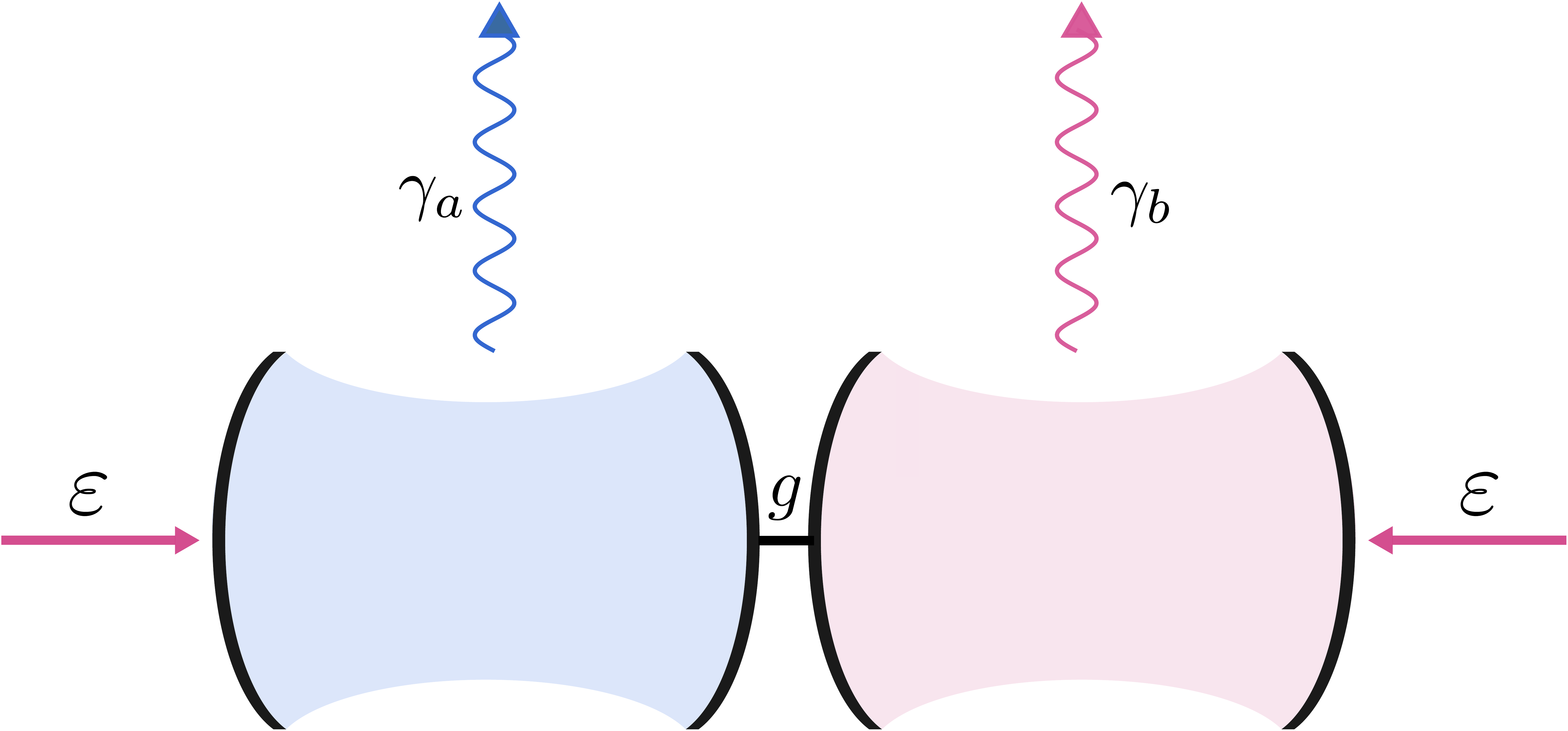}}
\caption{Schematic representation of the optical setup, in which the hidden ${\cal{PT}}$ symmetry is present.}
\label{fig:scheme}
\end{figure}

Let us apply the idea of EF to investigate the symmetry and EP of a non-Hermitian Hamiltonian of a physical system described by the master equation ($\hbar=1$)
\begin{eqnarray}    
\label{eq:masterEq}
\dot{\rho}=-i [H,\rho]
+\frac{1}{2}\sum_{i} \left(2 C_{i}\rho C^{+}_{i} 
- C^{+}_{i} C_{i}\rho - \rho C^{+}_{i} C_{i}\right) \, .
\end{eqnarray}
The Hamiltonian of the system is given by
\begin{eqnarray}
  \label{eq:Hamiltonian}
  H&=& g (a^{\dagger}b + b^{\dagger}a) +i\varepsilon (a-a^{\dagger}) +i\varepsilon (b-b^{\dagger})\, ,
\end{eqnarray}
and collapse operators are given by
\begin{eqnarray}
\label{eq:collapse_ops0}
C_1 = \sqrt{2\gamma_a}\,a\, , &\,&   C_2 = \sqrt{2\gamma_b}\,b\, .
\end{eqnarray}
Here, $g$ is a coupling strength, $a$ and $b$ denote the annihilation operators, $\gamma_{a}$, and $\gamma_{b}$ are the field damping rates of both modes. For simplicity, we assume that $g$ is real and positive. The above master equation describes a quantum system composed of two coupled optical cavities, which are both driven by a classical field, and from both of them a field leaks out to the reservoir, as shown in Fig.~\ref{fig:scheme}.
Therefore, this system contains both incoherent loss and coherent gain. This is a necessary condition to obtain a steady state solution different from the vacuum state. Such a system can serve as a source of light and it should be possible to implement it experimentally.

The master equation~(\ref{eq:masterEq}) can be rewritten to the form
\begin{eqnarray}    
\label{eq:masterEqNHH}
\dot{\rho}&=&-i (H_{\rm{_{nH}}}\rho-\rho H_{\rm{_{nH}}}^{\dagger}) 
+\sum_{i} C_{i}\rho C_{i}^{\dagger} \, ,
\end{eqnarray}
where
\begin{eqnarray}
\label{eq:generalHeff}
  H_{\rm{_{nH}}} = H - \frac{i}{2}\sum_i C_i^{\dagger}C_i\, .
\end{eqnarray}
Therefore, the non-Hermitian Hamiltonian is given by
\begin{eqnarray}
  \label{eq:HamiltonianNH0}
  H_{\rm{_{nH}}}&=& g (a^{\dagger}b + b^{\dagger}a) +i\varepsilon (a-a^{\dagger}) +i\varepsilon (b-b^{\dagger})\nonumber\\
  &&-i\gamma_{a} a^{\dagger}a - i\gamma_{b} b^{\dagger}b\, .
\end{eqnarray}
First, let us express Eq.~(\ref{eq:HamiltonianNH0}) in terms of the new bosonic operators defined by
\begin{eqnarray}
c=a +\varepsilon\alpha\hat{I}\, ,
&\quad& c^{+}=a^{+}+\varepsilon\beta \hat{I}\, ,\nonumber\\
d=b+\varepsilon\delta \hat{I}\, ,
&\quad& d^{+}=b^{+}+\varepsilon\theta \hat{I}\, ,
\end{eqnarray}
where $\alpha=(\gamma_{b}-i g)/\xi$, $\beta=-(\gamma_{b}-i g)/\xi$, $\delta=(\gamma_{a}-i g)/\xi$, $\theta=-(\gamma_{a}-i g)/\xi$ and $\xi=g^2+\gamma_{a}\gamma_{b}$.
Note that $c^{+}$($d^{+}$) is not Hermitian conjugation of $c$($d$). Hence, we have used the symbol "$+$" instead of "$\dagger$". Nevertheless, the operators $c$ and $d$ commute with each other and satisfy $[c,c^{+}]=1$ and $[d,d^{+}]=1$, so actually they satisfy commutation relations of independent oscillators. In terms of these new bosonic operators, the Hamiltonian takes the form
\begin{eqnarray}
  \label{eq:HamiltonianNH1}
  H_{\rm{_{nH}}}&=& g (c^{+}d + d^{+}c)-i\gamma_{a} c^{+}c - i\gamma_{b} d^{+}d-\chi \hat{I}\, ,
\end{eqnarray}
where $\chi=i 2\varepsilon^{2}\gamma/(g^2+\gamma^2-\kappa^2)$ and $\gamma=(\gamma_{a}+\gamma_{b})/2$. We have dropped the real part of $\chi$, because it contributes only an overall irrelevant phase factor. After introducing $\kappa=(\gamma_{a}-\gamma_{b})/2$, this Hamiltonian can be rewritten as a sum of two parts
\begin{eqnarray}
  \label{eq:HamiltonianPT05}
  H_{\rm{_{nH}}}&=&\underbrace{g (c^{+}d + d^{+}c)
  -i\kappa c^{+}c + i\kappa d^{+}d}_{=H^{\cal{PT}}}\nonumber\\
  &&\underbrace{- i\gamma(c^{+}c + d^{+}d) - \chi \hat{I}}_{=H_0}\, .
\end{eqnarray}
It can be verified that $H^{\cal{PT}}$ is ${\cal{PT}}$-symmetric using the spatial reflection defined by
\begin{eqnarray}
  \label{eq:def_P}
{\cal{P}}&=&P_{\rm{S}} \exp[i\pi(c^{\dagger}c+d^{\dagger}d)]\, ,
\end{eqnarray}
where $P_{\rm{S}}$ is the exchange operator~\cite{HorodeckiPRL02}, which spatially interchanges the modes (i.e. $c\leftrightarrow d$). A matrix representation of $P_{\rm{S}}$ is given by a perfect shuffle~\cite{Loan00}.
We define the time-reversal operator ${\cal{T}}$ just as the complex conjugation operator (${\cal{T}} i {\cal{T}}=-i$). 
Note that ${\cal{P}}$ given by Eq.~(\ref{eq:def_P}) is a reflection operator (i.e., ${\cal{P}}={\cal{P}}^{-1}$) and $[{\cal{P}},{\cal{T}}]=0$. Using it and formulas: $\exp(\alpha c^{\dagger} c) c \exp(-\alpha c^{\dagger} c) = \exp(-\alpha) c$ and $\exp(\alpha c^{\dagger} c) c^{\dagger} \exp(-\alpha c^{\dagger} c) = \exp(\alpha) c^{\dagger}$, one can easily check that $({\cal{PT}}) c ({\cal{PT}}) = -d$, $({\cal{PT}}) c^{\dagger}({\cal{PT}}) = -d^{\dagger}$, $({\cal{PT}}) d ({\cal{PT}}) = -c$, $({\cal{PT}}) d^{\dagger}({\cal{PT}}) = -c^{\dagger}$ and $({\cal{PT}}) i ({\cal{PT}}) = -i$.

To find the eigenvalues, we use bosonic algebra combined with Fock space representation of~(\ref{eq:HamiltonianNH1})~\cite{Teimourpour2018}. To this end, we introduce the operators $[e, f]^{\rm{T}}=\boldsymbol{R}\, [c, d]^{\rm{T}}$ and $[e^{+}, f^{+}]^{\rm{T}}=\boldsymbol{R}\, [c^{\dagger}, d^{\dagger}]^{\rm{T}}$,
where
\begin{equation}
  \label{eq:R}
\boldsymbol{R}\equiv\begin{bmatrix}
\cos\frac{\alpha}{2}&\sin\frac{\alpha}{2}\\
-\sin\frac{\alpha}{2}&\cos\frac{\alpha}{2}
\end{bmatrix}\, ,
\end{equation}
$\sin{(\alpha/2)}=\sqrt{(\Omega+i\kappa)/(2\Omega)}$, $\Omega=\sqrt{g^2-\kappa^2}$ and $\cos{(\alpha/2)}=\sqrt{(\Omega-i\kappa)/(2\Omega)}$.
The new operators satisfy the following commutation relations $[e,e^{+}]=1$, $[f,f^{+}]=1$, $[e,f^{+}]=0$, $[f,e^{+}]=0$, $[e,f]=0$ and $[f^{+},e^{+}]=0$, and therefore, can be considered as annihilation and creation operators~\cite{Teimourpour2018}.
In terms of these operators, the Hamiltonian takes the form
\begin{eqnarray}
  \label{eq:HamiltonianPT06}
  H_{\rm{_{nH}}}&=&\underbrace{\Omega\,(e^{+} e-f^{+} f)}_{=H^{\cal{PT}}}
  \underbrace{-i\gamma (e^{+} e+f^{+} f)- \chi \hat{I}}_{=H_0}\, .
\end{eqnarray}
Now, it is also easy to check that $[H^{\cal{PT}},H_{0}]=0$, so we can expect that $H_{\rm{_{nH}}}$ has a hidden ${\cal{PT}}$ symmetry. The geometric part of the Hamiltonian corresponds to its imaginary part~\cite{Zhang19a}, and therefore, assuming an unbroken ${\cal{PT}}$-symmetric phase, it is given just by $H_{0}$. This part is important because the rate at which each of the eigenstates scales in the EF is determined by the eigenvalues of $H_{0}$.
Note that the geometric part is not just the operator of the total number of photons in both modes. 

The eigenvalues of $H^{\cal{PT}}$ (EF) and $H_{\rm{_{nH}}}$ (IF) are given by
\begin{eqnarray}
  \label{eq:lPT}
  \lambda^{PT} &=& \Omega\, (N_{e} - N_{f})\, , \\
    \label{eq:lnH}
  \lambda^{\rm{nH}} &=&  \Omega\,(N_{e} - N_{f})-i\gamma\,(N_{e} + N_{f}) - \chi \hat{I}\, ,  
\end{eqnarray}
respectively. We have denoted excitation numbers in the supermodes $e$ and $f$ by $N_e$ and $N_f$. From these formulas, it is evident that all eigenvalues have the same value for $\kappa= g$ in EF. In IF all eigenvalues have the same real part but imaginary parts can be different. 
\begin{figure}[htbp]
\centering
\fbox{\includegraphics[width=0.7\linewidth]{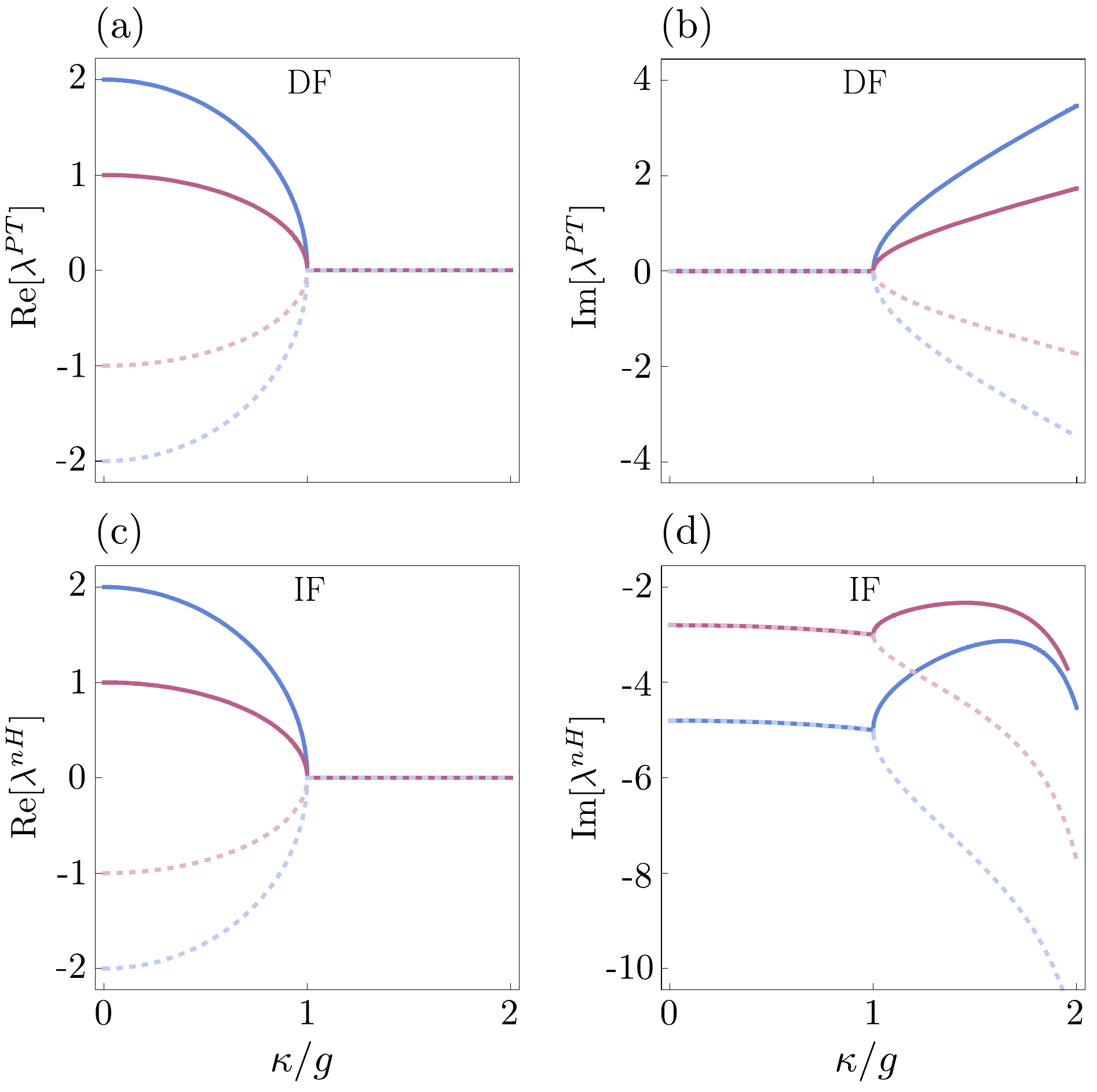}}
\caption{Eigenvalues of the Hamiltonian presented in the equilibrium frame (EF) given by Eq.~(\ref{eq:lPT}) [panels (a) and (b)] and in the initial frame (IF) given by Eqs.~(\ref{eq:lnH}) [panels (c) and (d)] as functions of the gain/loss coefficient $\kappa$, for $(\gamma,\varepsilon)/g=(2,1)$. Real parts are the same in both frames, but imaginary parts are different in these frames. Nevertheless, in both frames, the exceptional point exists and is placed at the same point of the parameter space, i.e., $g=\kappa$.}
\label{fig:passive-pt-ex2}
\end{figure}
To illustrate it, we have compared in Fig.~\ref{fig:passive-pt-ex2} eigenvalues corresponding to the following four eigenstates: $|\psi_1\rangle=|1\rangle_{e}|0\rangle_{f}$, $|\psi_2\rangle=|0\rangle_{e}|1\rangle_{f}$, $|\psi_3\rangle=|2\rangle_{e}|0\rangle_{f}$, and $|\psi_4\rangle=|0\rangle_{e}|2\rangle_{f}$ in both frames.
As expected for ${\cal{PT}}$-symmetric theory, it can be seen in panels (a) and (b) that in the EF all eigenvalues are real when a system is in unbroken, ${\cal{PT}}$-symmetric phase and complex-conjugate pairs of eigenvalues appear when ${\cal{PT}}$ symmetry is broken. It is also seen that a transition from broken to unbroken PT symmetry occurs at the point $\kappa= g$. This is EP, in which all four eigenvalues have the same real and imaginary parts. In this case, all eigenvalues at the EP are equal to zero. We can conclude from panels (a) and (b) that the investigated system is ${\cal{PT}}$-symmetric in EF and eigenvalues of the Hamiltonian describing this system behave exactly as predicted by the ${\cal{PT}}$-symmetric theory. 

Fig.~\ref{fig:passive-pt-ex2} in panels (c) and (d) show real and imaginary parts of the Hamiltonian~(\ref{eq:HamiltonianNH1}), which describes the system investigated in IF. In this frame all eigenvalues have always non-zero imaginary parts. Nevertheless, it is clearly seen that there is a correspondence between IF and EF. According to Eq.~(\ref{eq:lnH}), EP is also present in the point $\kappa= g$. In this case, however, the same real and imaginary parts have only such eigenvalues, which correspond to states with the same excitation number $N=N_{e} + N_{f}$. Therefore, there are two pairs of coalescing eigenvalues in panels (c) and (d): \{$\lambda^{\rm{nH}}_1$, $\lambda^{\rm{nH}}_2$\} and \{$\lambda^{\rm{nH}}_3$, $\lambda^{\rm{nH}}_4$\}. The first pair corresponds to $N=1$ and the second to $N=2$. 
It is also seen that for $\kappa<g$ eigenvalues corresponding to the same excitations number $N$ have different real parts and equal imaginary parts. For $\kappa>g$, these eigenvalues have different imaginary parts and equal real parts. Due to the similarities between EF and IF, we can say that $H_{\rm{_{nH}}}$ has a hidden ${\cal{PT}}$ symmetry despite the fact that $H_{\rm{_{nH}}}$ is not ${\cal{PT}}$-symmetric. Similarly, we can say that $\kappa<g$ is the region of unbroken ${\cal{PT}}$ hidden symmetry, and $\kappa>g$ is the region of broken ${\cal{PT}}$ hidden symmetry.

It is worth comparing the results presented in Fig.~\ref{fig:passive-pt-ex2} with the results presented in~\cite{Lange2020}. The real part of the eigenvalues in~\cite{Lange2020} diverge, whereas these seen in Fig.~\ref{fig:passive-pt-ex2} are finite and continuous. This is because the Hamiltonian~(\ref{eq:HamiltonianNH0}) describes a real physical system and the ${\cal{PT}}$-symmetric Hamiltonian considered in~\cite{Lange2020} is just a mathematical model. If we chose $\kappa>\gamma$, i.e., if we assumed an incoherent gain, then a divergence would also appear here.

Let us discuss the effect of the laser driving $\varepsilon$ on the eigenvalues. One can see that the laser driving changes only $\chi$, which is multiplied by an identity operator in the Hamiltonian. Since $\chi$ is purely imaginary, it increases equally the decay rates of all eigenstates. This leads to an increase in the probability of a collapse occurring during the observed time interval. Hence, driving the system by an external laser field increases the average rate at which photons are emitted from the system to the environment. 

We have shown that the non-Hermitian Hamiltonian $H_{\rm{_{nH}}}$ given by Eq.~(\ref{eq:HamiltonianNH0}) has a hidden ${\cal{PT}}$ symmetry and displays EP. Moreover, we know that $H_{\rm{_{nH}}}$ is appropriate for describing a real quantum system provided that the conditional evolution of the system is assumed. According to quantum trajectory theory~\cite{carmichael_traj,PlenioKnight_traj} the non-Hermitian Hamiltonian $H_{\rm{_{nH}}}$ describes the conditional time evolution of an open system when the system's interaction with the environment is monitored by perfect detectors. During the time intervals when no photon decay is detected, the evolution is governed by $H_{\rm{_{nH}}}$. This evolution is interrupted by collapses corresponding to the action of the collapse operators~(\ref{eq:collapse_ops0}). The quantum trajectory theory makes it possible to describe a state evolution conditioned on a sequence of detected collapses. The master equation approach cannot describe the state evolution conditioned on a particular detection record, because the master equation evolves all possible trajectories in time as a single package. In other words, the master equation approach does not assume any knowledge of the detection events.

The non-Hermitian Hamiltonian $H_{\rm{_{nH}}}$ corresponding to the master equation~(\ref{eq:masterEq}) gives us the opportunity to compare EP calculated from the non-Hermitian Hamiltonian (HEP) with EP calculated from a Liouvillian superoperator (LEP). In general, LEPs can be different from HEPs, because we calculate LEPs taking into account also the last term of the master equation~(\ref{eq:HamiltonianNH0}), i.e., $\sum_{i} C_{i}\rho C_{i}^{\dagger}$, which describes quantum jumps. $H_{\rm{_{nH}}}$ governs the evolution of the system in the absence of quantum jumps, and thus quantum jumps are not taken into account here. The best way to calculate LEP in the case of this infinite-dimensional system is to use the Heisenberg-Langevin equations averaged over the reservoir~\cite{Arkhipov_20b}. This approach is equivalent to calculating LEP from the Liouvillian superoperator as shown in Ref.~\citenum{perina2022quantum}. Knowing the Hamiltonian~(\ref{eq:Hamiltonian}) and the collapse operators~(\ref{eq:collapse_ops0}) we can obtain the Heisenberg-Langevin equation without noise terms for an operator $A$ using the following formula~\cite{Plankensteiner2022l}:
\begin{eqnarray}
\label{eq:HL_formula}
  \dot{A}= i [H,A] + \sum_k \frac{1}{2}\big(2 C_k^{\dagger} A C_k - C_k^{\dagger} C_k A - A C_k^{\dagger} C_k\big)\, .
\end{eqnarray}
In this way we obtain the closed set of differential equations for the fields' operator moments
\begin{eqnarray}
\label{eq:moments_eqs}
\langle \dot{a}\rangle&=&-\gamma_{a} \langle a\rangle -i g\langle b\rangle- \varepsilon\, ,\nonumber\\
\langle \dot{b}\rangle&=& -i g \langle a\rangle -\gamma_{b}\langle b\rangle- \varepsilon\, .
\end{eqnarray}
The matrix form of this set of linear equations is given by
\begin{eqnarray}    
    \boldsymbol{\dot{v}}&=& -i\boldsymbol{M}\,\boldsymbol{v} - \boldsymbol{v_0}\, ,
\end{eqnarray}
where
\begin{equation}
\boldsymbol{M} = \left(
\begin{array}{cc}
 -i\gamma_{a} & g  \\
 g  & -i\gamma_{b}  \\
\end{array}
\right)\, ,
\end{equation}
$\boldsymbol{v}=[\langle a\rangle, \langle b\rangle]^T$ and $\boldsymbol{v_0}=[\varepsilon,\varepsilon]^{\mathrm{T}}$. The diagonalisation of $\boldsymbol{M}$ leads to formulas for the eigenvalues $\lambda_{\pm}=\pm\Omega- i\gamma$ and the corresponding eigenvectors $\boldsymbol{v_{\pm}}=[\pm\Omega-i\kappa,\, g]^T$, where $\Omega=\sqrt{g^2-\kappa^2}.$ It is evident that the point $\kappa=g$ is LEP, because at this point both eigenvalues coincide and the corresponding eigenvectors coalesce. One can see that in the case of this optical system LEP is equivalent to HEP. 

\subsection{The Effect of Thermal Photons in the Reservoir on HEP in Circuit QED}
\begin{figure}[htbp]
\centering
\fbox{\includegraphics[width=0.7\linewidth]{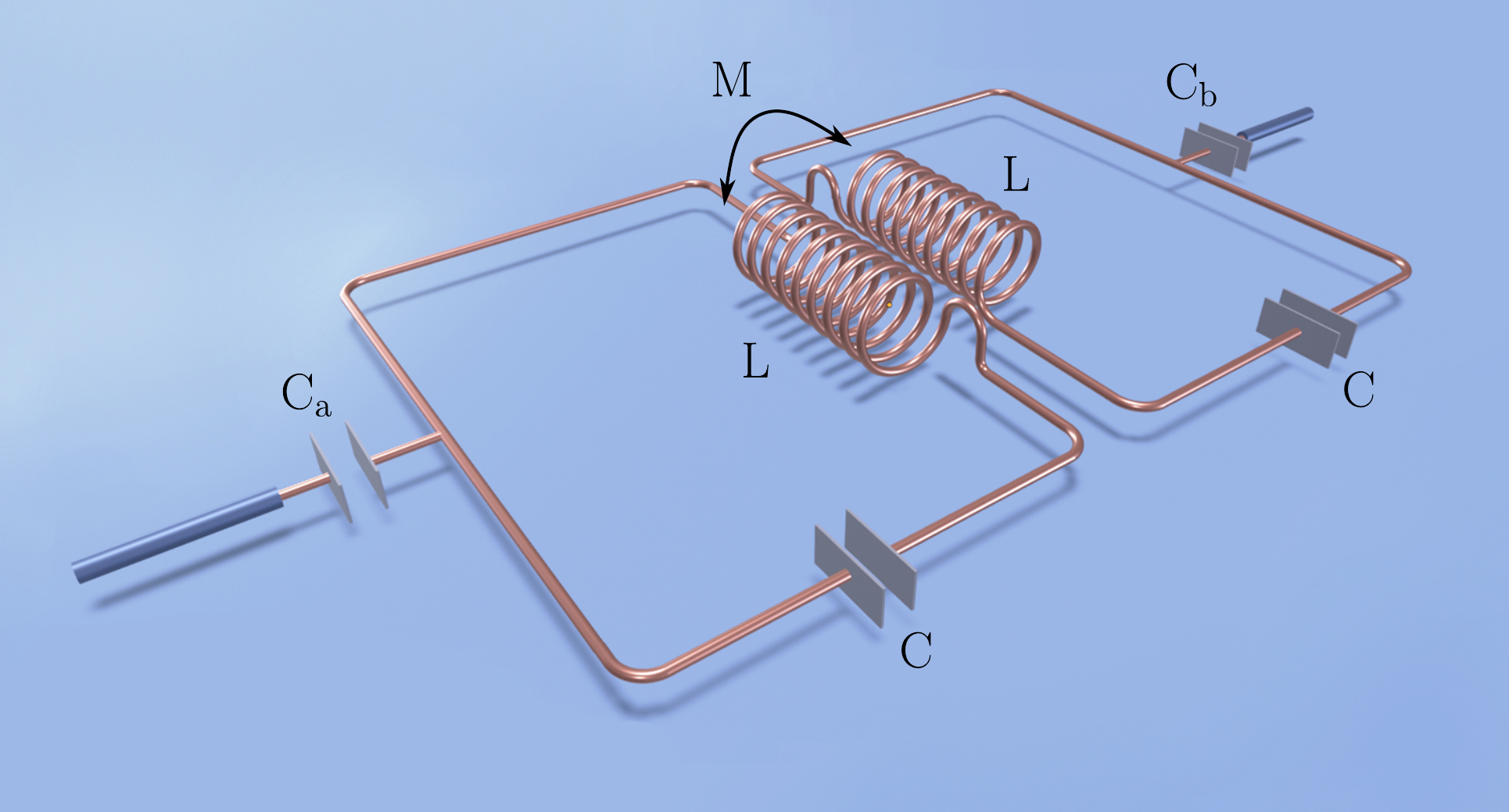}}
\caption{Schematic diagram of a superconducting circuit realised in the microwave domain, which is equivalent to the optical setup shown in Fig.~\ref{fig:scheme}. The difference between these setups is that the frequency of the optical cavity is four orders of magnitude higher than that of the superconducting resonator LC. Therefore, thermal photons present in the environment can be neglected in the optical case, but should be taken into account in the microwave case.}
\label{fig:circuit}
\end{figure}

Let us now consider another physical system --- two coupled superconducting resonators driven by an external electromagnetic field. This physical system should also be experimentally feasible~\cite{Purkayastha2020PRR}. The circuit diagram is shown in Fig.~\ref{fig:circuit}. 
These two LC resonators are inductively coupled to each other via mutual inductance $M$ and capacitively coupled to sources of loss via $\rm{C}_a$ and $\rm{C}_b$. These two capacitors, i.e., $\rm{C}_a$ and $\rm{C}_b$ also allow for driving the LC resonators. It is evident that this electrical circuit system can be also described by the Hamiltonian~(\ref{eq:Hamiltonian}). However, the master equation has to be modified because of thermal effects. In the case of optical systems, which interact with a thermal environment, the number of thermal photons is negligibly small. However, in the case of superconducting microwave resonators, even at liquid helium temperatures, a few thermal photons are present in the thermal bath~\cite{carmichael_traj}. It is easy to take into account the number of thermal photons by substituting the following collapse operators into the master equation~(\ref{eq:masterEq})
\begin{eqnarray}
\label{eq:collapse_ops_ther}
      C_1 = \sqrt{2\gamma_a (n+1)}a\, , &\,& 
      C_2 = \sqrt{2\gamma_a n} a^{\dagger}\, ,\nonumber \\
      C_3 = \sqrt{2\gamma_b (n+1)}b\, , &\,&
      C_4 = \sqrt{2\gamma_b n} b^{\dagger}\, .
\end{eqnarray}
For the sake of simplicity, we assume that both resonators are coupled to baths with the same number of thermal photons $n$, i.e., to baths with the same temperatures. Using Eq.~(\ref{eq:generalHeff}), we obtain the non-Hermitian Hamiltonian
\begin{eqnarray}
\label{eq:HNH24}
  H_{\rm{_{nH}}} &=& H - i\gamma_{a}(n+1) a^{\dagger} a - i\gamma_{a} n a a^{\dagger}- i\gamma_{b}(n+1) b^{\dagger} b \nonumber\\
  &&- i\gamma_{b} n b b^{\dagger}\, ,
\end{eqnarray}
which governs the evolution of the system between quantum jumps. Using the bosonic commutation relations we can rewrite the non-Hermitian Hamiltonian to the form
\begin{eqnarray}
\label{eq:HamiltonianNHtherm}
H_{\rm{_{nH}}} &=& g (a^{\dagger}b + b^{\dagger}a) +i\varepsilon (a-a^{\dagger}) +i\varepsilon (b-b^{\dagger}) \nonumber\\
&&- i\gamma'_{a} a^{\dagger} a - i\gamma'_{b} b^{\dagger} b- \chi_{t}\hat{I}\, ,
\end{eqnarray}
where $\gamma'_{a}=\gamma_{a}( 2 n+1)$, $\gamma'_{b}=\gamma_{b} (2 n +1)$ and $\chi_{t}=i\, n(\gamma_{a} +\gamma_{b})$. Note that the Hamiltonians given by Eq.~(\ref{eq:HamiltonianNHtherm}) and Eq.~(\ref{eq:HamiltonianNH0}) have a very similar form. Hamiltonian~(\ref{eq:HamiltonianNHtherm}) has a different damping constants and an extra term --- an identity operator multiplied by a constant. Therefore, this microwave system also has hidden ${\cal{PT}}$ symmetry and the point in which a transition from broken to unbroken hidden ${\cal{PT}}$ symmetry takes place. 
Due to this similarity, we can diagonalise non-Hermitian Hamiltonian~(\ref{eq:HamiltonianNHtherm}) in exactly the same way as Hamiltonian~(\ref{eq:HamiltonianNH0}), which yields
\begin{eqnarray}
  \label{eq:H_diag_therm}
  H_{\rm{_{nH}}}&=& \Omega'\,(e^{+} e-f^{+} f)-i\gamma' (e^{+} e+f^{+} f)- \chi' \hat{I}\, ,
\end{eqnarray}
where 
$\Omega'=\sqrt{g^2-\kappa^{\prime\,2}}$, $\gamma'=(2 n +1)\gamma$, $\kappa'=(2 n +1)\kappa$ and
\begin{eqnarray}
\chi'&=&i\,2\gamma\Big[n+\frac{\varepsilon^{2} (2 n +1)}{g^2+(\gamma^2 -\kappa^2)(2 n +1)^2}\Big]\, .
\end{eqnarray}
The eigenvalues of this microwave system are then given by
\begin{eqnarray}
  \label{eq:lnHthem}
  \lambda^{\rm{nH}} &=&  \Omega'\,(N_{e} - N_{f})-i\gamma'\,(N_{e} + N_{f}) - \chi' \hat{I}\, .
\end{eqnarray}
\begin{figure}[htbp]
\centering
\fbox{\includegraphics[width=0.9\linewidth]{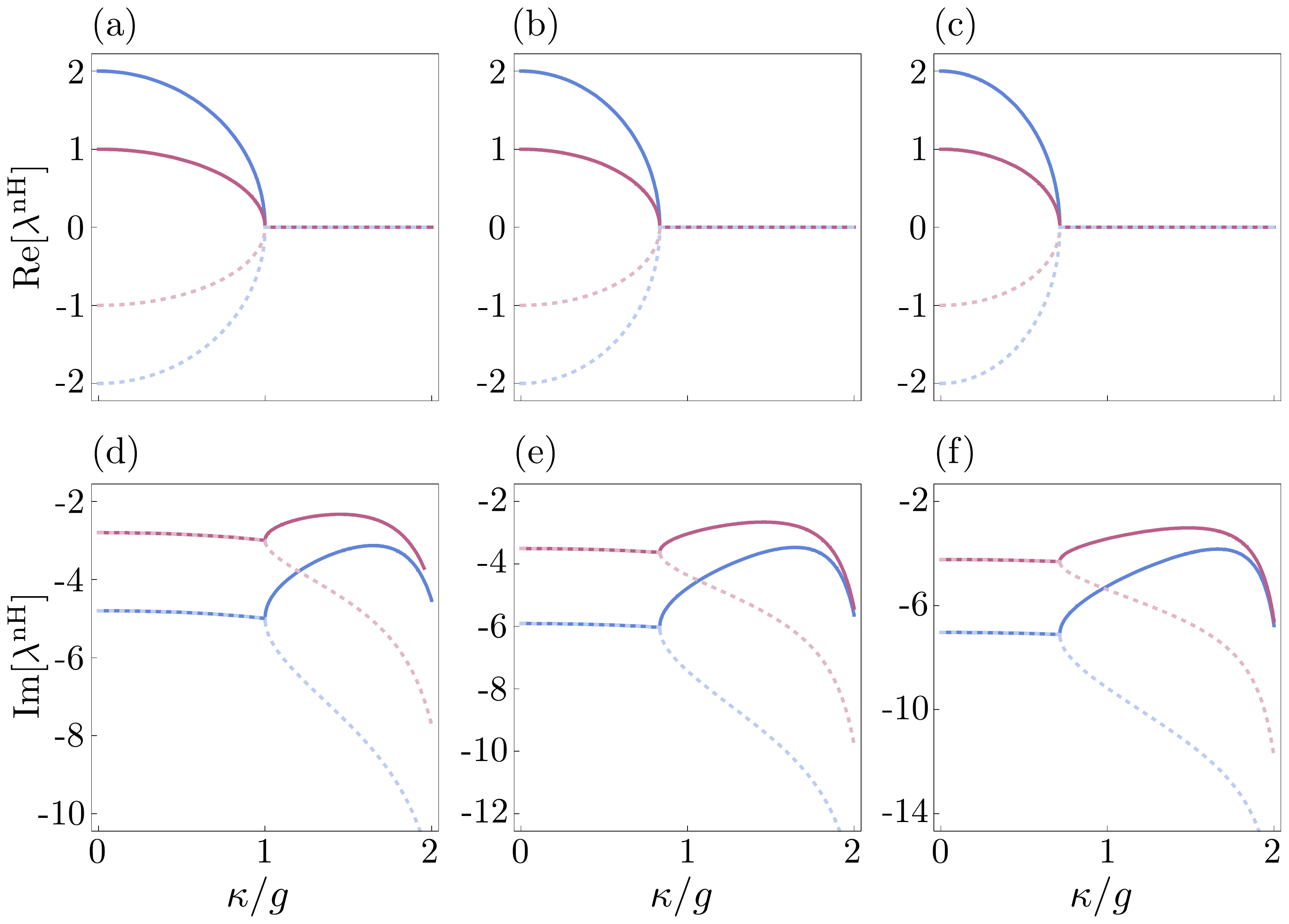}}
\caption{Real and imaginary parts of eigenvalues given by Eq.~(\ref{eq:lnHthem}) as functions of $\kappa/g$ for different number of thermal photons. Parameters: $\gamma$ = 2, $g$ = 1, $\varepsilon$ = 1. Panels (a) and (d): $n$ = 0, panels (b) and (e): $n$ = 0.1, panels (c) and (f): $n$ = 0.2.}
\label{fig:thermal}
\end{figure}
In Fig.~\ref{fig:thermal} we have plotted for different values of $n$ four eigenvalues corresponding to the four eigenstates, which were considered in the case of the optical system, i.e., $|\psi_1\rangle$, $|\psi_2\rangle$, $|\psi_3\rangle$, and $|\psi_4\rangle$. It can be seen that the system has a hidden ${\cal{PT}}$ symmetry and displays HEP between the unbroken and broken phases. It can also be seen that the position of HEP is dependent on the number of thermal photons $n$. From Eq.~(\ref{eq:lnHthem}) we can infer that the eigenvalues corresponding to the eigenstates with the same number of excitations are equal for $g$ satisfying the condition:
\begin{eqnarray}
  g_{_{\rm{HEP}}} &=&(2 n +1)\kappa \, . 
\end{eqnarray}
Therefore, the effect of thermal photons in the reservoir on the HEP is to move it in such a direction that the parametric region of the unbroken hidden ${\cal{PT}}$ symmetry is reduced.

Let us now calculate the eigenvalues of Liouvillian using Heisenberg-Langevin equations averaged over the reservoir to obtain the formula for LEP. In order to generate equations for the fields' operator moments we substitute the Hamiltonian~(\ref{eq:Hamiltonian}) and collapse operators~(\ref{eq:collapse_ops_ther}) into Eq.~(\ref{eq:HL_formula}). We obtain the same set of differential equations as in the case of the optical setup, i.e., the set given by Eqs.~(\ref{eq:moments_eqs}). So, we obtain also the same eigenfrequencies. Thus, the system displays LEP for g satisfying the following condition:
\begin{eqnarray}
  g_{_{\rm LEP}} &=&\kappa \, . 
\end{eqnarray}
It should be noted that $g_{_{\rm LEP}}$ is independent of $n$ in contrast to $g_{_{\rm HEP}}$. Therefore, the effect of thermal photons is that LEP is not equivalent to HEP in this circuit QED system. This effect of thermal photons on EPs has not been observed yet, to the best of our knowledge. 

This circuit QED system should be experimentally feasible, so it should also be possible to observe the spectrum of the field radiated by this system. Therefore, the question arises whether this circuit QED system displays EP in the point $g_{_{\rm LEP}}$ or $g_{_{\rm HEP}}$. As mentioned earlier, a non-Hermitian Hamiltonian in the quantum trajectory method describes a conditional evolution of a system, which is monitored by perfect detectors, whereas the master equation method describes an evolution of an open system, for which a sequence of jumps events is not known. We note that HEPs are observed in experiments that include conditional measurements~\cite{Minganti_20}.

Finally, let us explain the difference in the positions of EPs
revealed by considering the non-Hermitian
Hamiltonian $H_{\rm nH}$~(\ref{eq:HNH24}) and by analysing the
dynamical matrix of the equations~(\ref{eq:moments_eqs}) for mean values
\cite{perina2022quantum}. The Hamiltonian $H_{\rm nH} $ in Eq.~(\ref{eq:HNH24}) describes two bosonic modes interacting with two independent
reservoirs at finite temperature, i.e., with non-zero mean reservoir
photon numbers $n$. Without the loss of generality, let us
concentrate our attention to one bosonic mode. The form of the
non-Hermitian Hamiltonian $H_{\rm nH}$ corresponds to the
following master equation for the mode statistical operator $\rho$ describing its interaction with the reservoir
\cite{Perina1991}:
\begin{eqnarray}
 \dot{\rho} &=& \gamma_a (n+1) \left(
  [a\rho,a^\dagger] +
  [a,\rho a^\dagger]\right) \nonumber \\
  & & + \gamma_a n \left(
  [a^\dagger\rho,a] +
  [a^\dagger,\rho a]\right).
 \label{eq:A1}
\end{eqnarray}

To identify the drift and diffusion terms in the evolution of the
mode as described by the master equation~(\ref{eq:A1}), let us rewrite it
for the quasi-distribution function $\Phi_{\cal N}$ introduced
in the Glauber-Sudarshan representation of the statistical
operator $\rho$ in the basis of coherent states $|\alpha\rangle$~\cite{Perina1991}:
\begin{equation}
 \rho = \int d^2\alpha \, \Phi_{\cal N}(\alpha,\alpha^*)
  |\alpha\rangle \langle \alpha|.
 \label{eq:A2}
\end{equation}
Using the properties of coherent states, we arrive at the
following Fokker-Planck equation \cite{Risken1996}:
\begin{eqnarray}
 \frac{d \Phi_{\cal N} }{dt} = \gamma_a
 \frac{\partial}{\partial \alpha} (\alpha \Phi_{\cal N}) +
 \gamma_a  \frac{\partial}{\partial \alpha^*} (\alpha^* \Phi_{\cal N}).
 \label{eq:A3}
\end{eqnarray}
According to Eq.~(\ref{eq:A3}), the drift terms correspond to the following
Heisenberg-Langevin equations for the operators $a$ and $a^\dagger$,
\begin{equation}
  \dot{a} = -\gamma_a a + L,
  \hspace{5mm}   \dot{a}^\dagger = -\gamma_a a^\dagger +L^\dagger,
\label{eq:A4}
\end{equation}
and the stochastic operator forces $L$ and $L^\dagger$ serve to 
describe the influence of the diffusion
term. The form of Eq.~(\ref{eq:A4}) corresponds to the following
non-Hermitian Hamiltonian $H_{\rm nH}^{\rm drift}$,
\begin{equation}
  H_{\rm nH}^{\rm drift} = -i \gamma_a a^\dagger a,
\label{eq:A5}
\end{equation}
that completely describes the mode evolution caused by the drift terms.

The consideration of non-Hermitian Hamiltonian~(\ref{eq:HNH24}) in the form of
Eq.~(\ref{eq:A5}), i.e.
\begin{equation}
  H_{\rm nH}^{\rm drift} = H -i\gamma_a a^\dagger a - i\gamma_b b^\dagger b
  \label{eq:A6}
\end{equation}
then leads to the EPs according to the condition (30) that
identifies LEPs.

The approach that gives the non-Hermitian
Hamiltonian of Eq.~(\ref{eq:HNH24}) incorporates the drift terms only partly,
which results in shifted temperature-dependent positions of EPs.
Subsequent inclusion of quantum jumps \cite{Minganti_20} then has
to correct for both the dynamics of the drift terms and the
diffusion terms. Once this correction is done, the positions of
EPs (HEPs) shift to those identified from the analysis of the
whole Liouvillian of the system (LEPs).

\section{Discussion}
We have introduced the concept of the equilibrium frame (EF) --- a frame, which scale is time dependent. This frame makes it possible to reveal ${\cal{PT}}$ symmetry hidden in passive non-Hermitian Hamiltonians, which do not have an incoherent gain term. More specifically, we have shown that if a non-${\cal{PT}}$-symmetric non-Hermitian Hamiltonian can be expressed as a sum of two terms (a ${\cal{PT}}$-symmetric term and a second term commuting with the first one), then such Hamiltonian has a hidden ${\cal{PT}}$ symmetry. Using the EF method we have proved that the non-Hermitian Hamiltonian describing a quantum system composed of two coupled optical cavities, which are both driven by a classical field, and from both of them a field leaks out to the reservoir, has a hidden ${\cal{PT}}$ symmetry. Although systems composed of two coupled optical resonators have been investigated many times in the context of ${\cal{PT}}$ symmetry~\cite{Teimourpour2018,Arkhipov_20b,Lange2020}, to our knowledge, the non-Hermitian Hamiltonian~(\ref{eq:HamiltonianNH0}) with hidden ${\cal{PT}}$ symmetry and coherent gain has been never presented. The presence of the coherent gain is important because it allows for using this system as a source of light. Thanks to the EF method, we have identified the region of the parameter space where the hidden ${\cal{PT}}$ symmetry is unbroken and the region of the broken phase. In this way, we have also found the position of the exceptional point (EP), where a transition from the broken phase to the unbroken phase takes place.

The non-Hermitian Hamiltonian~(\ref{eq:HamiltonianNH0}) describes a conditional evolution of a feasible optical open system interacting with its environment. Therefore, it is possible to write the master equation for the considered system. We have written the master equation that describes the evolution of this optical system and we have calculated EP from a Liouvillian superoperator. We have found that in the case of this optical system, EP obtained from the Liouvillian superoperator (LEP) is equivalent to EP determined from the non-Hermitian Hamiltonian (HEP). Next, we have investigated a circuit QED system, which is described by the same Hermitian Hamiltonian as the optical system. However, in this microwave system, the effect of thermal photons present in the environment cannot be neglected, so the master equation is different than in the optical case. We have found that in the case of the circuit QED system LEPs differ from HEPs as a consequence of non-zero number of thermal photons. This means breaking the equivalence between HEPs and LEPs.

\begin{acknowledgments}
This work was supported by the Polish National Science Centre (NCN) under the Maestro Grant No. DEC-2019/34/A/ST2/00081. KB also acknowledges financial support from the  Czech Science Foundation under project No. 19-19002S and project No. CZ.1.05/2.1.00/19.0377 of the Ministry of Education, Youth and Sports of the Czech Republic. JP acknowledges the support from M\v{S}MT \v{C}R projects No. CZ.02.2.69/0.0/0.0/18\_053/0016919.
\end{acknowledgments}

%

\end{document}